\documentstyle[11pt,AATS,epsf]{article}	

\begin{document}	

\title{R-process pattern in the Very-Metal-Poor Halo Star CS 31082-001} 

\author{Vanessa Hill}
\affil{European Southern Observatory, Germany}
\author{Bertrand Plez}
\affil{GRAAL, Universit\'e de Montpellier 2, France}
\author{Roger Cayrel}
\affil{DASGAL, Observatoire de Paris, France}
\author{Timothy C. Beers}
\affil{Michigan State University, USA}


\begin{abstract}
The very-metal-poor halo star CS\,31082-001
was discovered to be 
very strongly r-process-enhanced during the course of a
VLT+UVES high-resolution follow-up of metal-poor
stars identified in the HK survey of Beers \& colleagues.
Both the strong n-capture element enhancement and
the low carbon and nitrogen content of the star (reducing the CN 
molecular band contamination) led to the first $^{238}$U
abundance measurement in a stellar spectrum (Cayrel et al. 2001), and
the opportunity to use both radioactive species $^{238}$U and
$^{232}$Th for dating the progenitor to this star. 
However, age computations all rely on the hypothesis that the
r-process pattern is solar, as this was indeed observed in the
other famous r-process-enhanced very metal poor stars CS\,22892-052
(Sneden et al. 1996, 2000) and in HD\,115444 (Westin et al. 2000).
Here, we investigate whether this hypothesis is verified also for CS\,31082-001, 
using a preliminary analysis of over 20 abundances
of n-capture elements in the range Z=38 to Z=92.
\end{abstract}

Cayrel et al. (2001; this volume) discuss the discovery and importance of CS
31082-001, and here we report a summary of a preliminary abundance analysis for
this star (Table \ref{table}).

\begin{figure}[htbp]
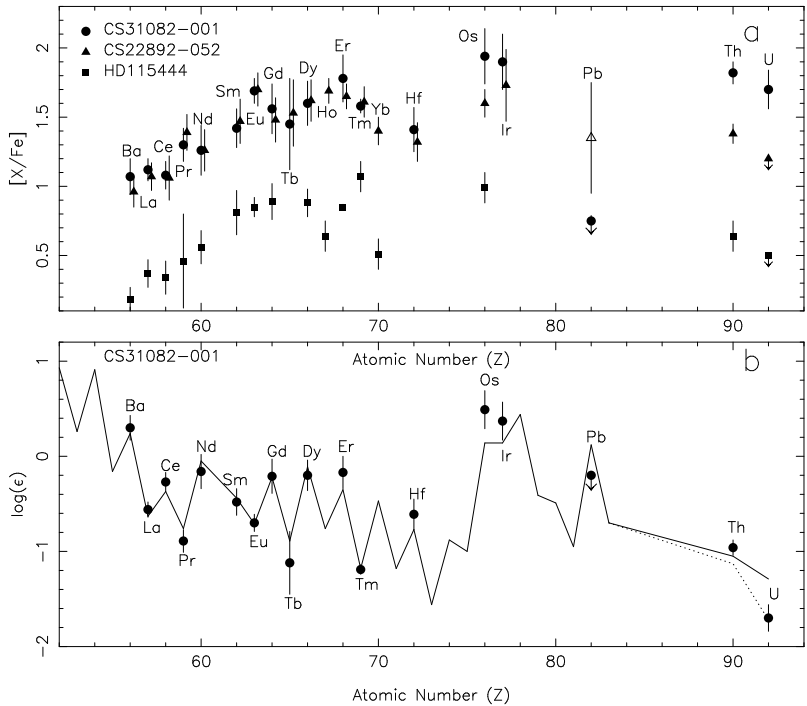

\plotfiddle{hill1.ps}{4.1cm}{-90}{45}{45}{-170}{250}
\plotfiddle{hill2.ps}{4.cm}{-90}{45}{45}{-170}{250}
\caption{N-capture abundances pattern in CS\,31082-001 compared to: 
{\it a-} CS\,22892-052 and HD\,115444. 
(The abscissa for CS\,22892-052 has been 
artificially shifted by +0.3 for readability).
{\it b-} to the solar
system r-process (Burris et al. 2000), scaled to match the
56$\geq$Z$\leq$72 abundances of CS\,31082-001. The radioactive
species (Th and U) solar system abundances are the values at the time
of formation of the solar system. The dotted line
show the abundances observed {\it today} for these two species.}\label{fig-cs3122}
\end{figure}

The n-capture element abundances (relative to iron, [X/Fe]) of CS\,31082-001 are
compared in Fig. \ref{fig-cs3122}-a to those of CS\,22892-052
and HD\,115444, showing that  
the overabundance of the Z$>$56 n-capture elements
in CS\,31082-001 is almost identical to that of CS\,22892-052, with a mean
overabundance of [X/Fe]$\sim$+1.7dex. These two stars are therefore the most
extreme cases of n-capture element enhancement in halo stars, far more
extreme than HD\,115444.
Furthermore, the abundance pattern of the 56$<$Z$<$70 elements in CS\,31082-001 are
indistinguishable from that of CS\,22892-052 or HD\,115444.
In contrast, the abundance pattern of  Z$>$70 seems to be 
more abundant in CS\,31082-001 than in CS\,22892-052 or
HD\,115444, including thorium, which is a factor four more 
abundant in CS\,31082-001 than in CS\,22892-052. Therefore the $\log\epsilon$(Th/Eu) ratio,
often used as an age indicator, is a factor 3 larger in CS\,31082-001
than in CS\,22892-052.

\begin{table}		
\begin{center}		
\caption{Neutron-capture elements abundances in CS\,31082-001. \label{table} }
\begin{tabular}{lcccc|lcccc}
El.& Z   &$log \epsilon$& $\sigma$& N$_{lines}$&El.& Z   &$log \epsilon$& $\sigma$& N$_{lines}$\\
\tableline 
Sr &38   & 0.68   &0.09   &  4 & Tb &65   &-1.12   &0.33   &  7 \\ 
 Y &39   &-0.16   &0.11   &  9 & Dy &66   &-0.20   &0.16   &  7 \\
Zr &40   & 0.47   &0.13   &  5 & Er &68   &-0.17   &0.17   &  5 \\
Ba &56   & 0.30   &0.13   &  7 & Tm &69   &-1.19   &0.05   &  3 \\
La &57   &-0.56   &0.08   &  4 & Hf &72   &-0.61   &0.16   &  2 \\
Ce &58   &-0.27   &0.10   &  9 & Os &76   & 0.49   &0.20   &  3 \\ 
Pr &59   &-0.89   &0.12   &  4 & Ir &77   & 0.37   &0.2    &  1 \\ 
Nd &60   &-0.16   &0.18   & 17 & Pb &82   &$<$-0.2:&       &  1 \\
Sm &62   &-0.48   &0.14   &  9 & Th &90   &-0.96   &0.08   & 11 \\
Eu &63   &-0.70   &0.09   &  9 &  U &92   &-1.70   &0.14   &  1 \\
Gd &64   &-0.21   &0.18   &  7 \\
\tableline \tableline
\end{tabular}

T$_{eff}=4825$K $\log g=1.5$
$\xi_{micro}=1.8{\rm km s^{-1}}$ [Fe/H]=-2.9
\end{center}
\end{table}

In Fig. \ref{fig-cs3122}-b, the abundance pattern of the n-capture
elements of CS\,31082-001 are compared to solar system r-process
pattern (as of Burris et al. 2000) scaled to match the mean
56$\geq$Z$\leq$72 n-capture elements abundance of CS\,31082-001
$\log\epsilon_{CS\,31082-001}-\log\epsilon_{SS}$=-1.22$\pm$0.03
($\sigma$=0.10 over 13 elements). 
Here again, whereas  the 56$<$Z$<$70 elements in
CS\,31082-001 are very well reproduced by a solar r-process, 
the Z$>$70 elements
are behaving in a somewhat more erratic way. While Os and Ir seem to be
more abundant than the scaled solar r-process, Pb
is notably underabundant (even the strongest line at 4057\AA\,
was not detected). 

This has the very interesting consequence that
the [Th/Eu] ratio (Eu or any other 56$\leq$Z$\leq$70 element) would
predict an epoch of formation of the 
n-capture elements present in CS\,31082-001 {\it later} than the epoch of
formation of the n-capture elements which enriched the solar system !
This conflicts with the observed U/Th ratios observed in CS\,31082-001 and
the solar system: $^{238}$U has a half-life a factor 3 shorter than
$^{232}$Th, so if the r-process elements of CS\,31082-001 were produced
after those of the solar system, the U/Th would be
significantly smaller in CS\,31082-001 than in the
solar-system (dotted line), which is not observed.
In fact, the age of CS\,31082-001 predicted from the [Th/Eu] ratio conflicts
with those from the [U/Th] [U/Os] or [U/Ir] ratios.

Beyond the issue of the age of this particular star, the fact
that the Z$>$70 elements pattern does not seem to be well-matched
by those of other similar stars (CS\,22892-052, HD\,115444) nor the solar-system
r-process elements is worrisome concerning the used of Th/Eu (or U/Eu)
ratios as age-tracers. The normalization of radioactive elements
abundances to elements {\em in the same
mass-range} becomes indispensable. 

The reason for the discrepancy of the Z$>$70 elements could be a
direct consequence of chemical inhomogeneities in the early Galaxy:
the ISM giving birth to very metal poor stars has probably only been
polluted by a 
very limited number of supernovae, and hence it is possible
that we now see the various outcomes of single events. Only
{\it significant samples} of such n-capture enhanced
elements will give clues to this issue.
Christlieb et al. (this volume)
suggest one method for quickly achieving this goal.

\end{document}